\documentclass[11pt]{article}
\usepackage{mathrsfs}
\usepackage[T1]{fontenc}
\usepackage{mathpazo}
\usepackage{setspace}
\usepackage{amsfonts}
\usepackage{amssymb}
\usepackage{epsfig}
\usepackage{latexsym}
\usepackage{color}
\usepackage{graphicx}
\usepackage{nicefrac}
\usepackage[latin1]{inputenc}
\usepackage{slashed}
\usepackage{multirow}

\usepackage{hyperref}
\def\hybrid{\topmargin -30pt    \oddsidemargin 0pt %%%%%%%%%%%%%% Archive-30pt
        \headheight 0pt \headsep 0pt
        \textwidth 6.25in       % A4 paper
        \textheight 9.5in       % A4 paper
        \marginparwidth .875in
        \parskip 5pt plus 1pt   \jot = 1.5ex}

\hybrid

\def\baselinestretch{1.2}

\catcode`\@=11

\def\marginnote#1{}
\newcount\hour
\newcount\minute
\newtoks\amorpm
\hour=\time\divide\hour by60
\minute=\time{\multiply\hour by60 \global\advance\minute by-\hour}
\edef\standardtime{{\ifnum\hour<12 \global\amorpm={am}%
        \else\global\amorpm={pm}\advance\hour by-12 \fi
        \ifnum\hour=0 \hour=12 \fi
        \number\hour:\ifnum\minute<10 0\fi\number\minute\the\amorpm}}
\edef\militarytime{\number\hour:\ifnum\minute<10 0\fi\number\minute}
\def\draftlabel#1{{\@bsphack\if@filesw {\let\thepage\relax
   \xdef\@gtempa{\write\@auxout{\string
      \newlabel{#1}{{\@currentlabel}{\thepage}}}}}\@gtempa
   \if@nobreak \ifvmode\nobreak\fi\fi\fi\@esphack}
        \gdef\@eqnlabel{#1}}
\def\@eqnlabel{}
\def\@vacuum{}
\def\draftmarginnote#1{\marginpar{\raggedright\scriptsize\tt#1}}

\def\draft{\oddsidemargin -.5truein
        \def\@oddfoot{\sl preliminary draft \hfil
        \rm\thepage\hfil\sl\today\quad\militarytime}
        \let\@evenfoot\@oddfoot \overfullrule 3pt
        \let\label=\draftlabel
        \let\marginnote=\draftmarginnote
   \def\@eqnnum{(\theequation)\rlap{\kern\marginparsep\tt\@eqnlabel}%
\global\let\@eqnlabel\@vacuum}  }

\def\draft2{
        \def\@oddfoot{\sl preliminary draft \hfil
        \rm\thepage\hfil\sl\today\quad\militarytime}
        \let\@evenfoot\@oddfoot \overfullrule 3pt
        \let\label=\draftlabel
        \let\marginnote=\draftmarginnote
   \def\@eqnnum{(\theequation)\rlap{\kern\marginparsep\tt\@eqnlabel}%
\global\let\@eqnlabel\@vacuum}  }

\def\preprint{\twocolumn\sloppy\flushbottom\parindent 2em
        \leftmargini 2em\leftmarginv .5em\leftmarginvi .5em
        \oddsidemargin -.5in    \evensidemargin -.5in
        \columnsep .4in \footheight 0pt
        \textwidth 10.in        \topmargin  -.4in
        \headheight 12pt \topskip .4in
        \textheight 6.9in \footskip 0pt
        \def\@oddhead{\thepage\hfil\addtocounter{page}{1}\thepage}
        \let\@evenhead\@oddhead \def\@oddfoot{} \def\@evenfoot{} }

\def\numberbysection{\@addtoreset{equation}{section}
        \def\theequation{\thesection.\arabic{equation}}}

\def\underline#1{\relax\ifmmode\@@underline#1\else
        $\@@underline{\hbox{#1}}$\relax\fi}

\def\titlepage{\@restonecolfalse\if@twocolumn\@restonecoltrue\onecolumn
     \else \newpage \fi \thispagestyle{empty}\c@page\z@
        \def\thefootnote{\fnsymbol{footnote}} }

\def\endtitlepage{\if@restonecol\twocolumn \else \newpage \fi
        \def\thefootnote{\arabic{footnote}}
        \setcounter{footnote}{0}}  %\c@footnote\z@ }

\catcode`@=12
\relax

\def\figcap{\section*{Figure Captions\markboth
        {FIGURECAPTIONS}{FIGURECAPTIONS}}\list
        {Figure \arabic{enumi}:\hfill}{\settowidth\labelwidth{Figure
999:}
        \leftmargin\labelwidth
        \advance\leftmargin\labelsep\usecounter{enumi}}}
 \relax
\def\tablecap{\section*{Table Captions\markboth
        {TABLECAPTIONS}{TABLECAPTIONS}}\list
        {Table \arabic{enumi}:\hfill}{\settowidth\labelwidth{Table
999:}
        \leftmargin\labelwidth
        \advance\leftmargin\labelsep\usecounter{enumi}}}
 \relax
\def\reflist{\section*{References\markboth
        {REFLIST}{REFLIST}}\list
        {[\arabic{enumi}]\hfill}{\settowidth\labelwidth{[999]}
        \leftmargin\labelwidth
        \advance\leftmargin\labelsep\usecounter{enumi}}}
 \relax
\makeatletter
\newcounter{pubctr}
\def\publist{\@ifnextchar[{\@publist}{\@@publist}}
\def\@publist[#1]{\list
        {[\arabic{pubctr}]\hfill}{\settowidth\labelwidth{[999]}
        \leftmargin\labelwidth
        \advance\leftmargin\labelsep
        \@nmbrlisttrue\def\@listctr{pubctr}
        \setcounter{pubctr}{#1}\addtocounter{pubctr}{-1}}}
\def\@@publist{\list
        {[\arabic{pubctr}]\hfill}{\settowidth\labelwidth{[999]}
        \leftmargin\labelwidth
        \advance\leftmargin\labelsep
        \@nmbrlisttrue\def\@listctr{pubctr}}}
 \relax
\makeatother

\def\be{\begin{equation}}
\def\ee{\end{equation}}
\def\ba{\begin{eqnarray}}
\def\ea{\end{eqnarray}}

\def\del{\partial}

\def\G{\Gamma}

\def\D{\Delta}

\def\th{\theta}

\def\m{\mu}
\def\n{\nu}
\def\om{\omega}

\def\l{\lambda}

\def\s{\sigma}
\def\S{\Sigma}

\def\cN{{\cal N}}

\def\no{\noindent}

\def\qq{\qquad}

\def\IR{\relax{\rm I\kern-.18em R}}

\def\inv{^{\raise.0ex\hbox{${\scriptscriptstyle -}$}\kern-.05em 1}}

\def \ha {{\frac{1}{2}}}

\def \ov {\over}

\begin{document}
%\draft2

\renewcommand{\theequation}{\arabic{equation}}
\renewcommand{\theequation}{\thesection.\arabic{equation}}

\renewcommand{\theequation}{\thesection.\arabic{equation}}
\csname @addtoreset\endcsname{equation}{section}

\begin{titlepage}
\begin{center}

{}\hfill DMUS--MP--12/11

\phantom{xx}
\vskip 0.5in

{\large \bf On Non-Abelian T-Duality and new ${\cal N}=1$ backgrounds}

\vskip 0.45in

  {\bf Georgios Itsios}${}^{1,3}\,$\footnote{{\tt gitsios@upatras.gr} }, {\bf Carlos N\'u\~nez}${}^{2}\,$\footnote{{\tt c.nunez@swansea.ac.uk} },\\  {\bf Konstadinos Sfetsos}${}^{3,1}\,$\footnote{{\tt k.sfetsos@surrey.ac.uk}}\ and\ {\bf Daniel~C.~Thompson}${}^{4}$\footnote{{\tt dthompson@tena4.vub.ac.be}}

\vskip 0.2in

${}^1$Department of Engineering Sciences, University of Patras,\\
26110 Patras, Greece\\
 \vskip .1in
${}^2$Swansea University, School of Physical Sciences,\\
Singleton Park, Swansea, SA2 8PP,  UK\\
\vskip .1in
${}^3$Department of Mathematics, University of Surrey,\\
 Guildford GU2 7XH, UK\\
\vskip .1in
${}^4$  Theoretische Natuurkunde, Vrije Universiteit Brussel, and \\
The International Solvay Institutes\\
Pleinlaan 2, B-1050, Brussels, Belgium \\

 \vskip .2in
\end{center}

\vskip .4in

\centerline{\bf Abstract}

\no
We study the action of non-Abelian T-duality in the context of ${\cal N}=1 $ geometries with
well understood field theory duals. In the conformal case this gives rise to a new solution that contains an $AdS_5 \times S^2$ piece.
In the case of non-conformal geometries we obtain a new background in massive IIA supergravity that presents similar
behaviour to the cascade of Seiberg dualities. Some physical observables are discussed.

\end{titlepage}
\vfill
\eject

\def\baselinestretch{1.2}
\baselineskip 10 pt
\noindent

%\tableofcontents

\def\baselinestretch{1.2}
\baselineskip 20 pt
\no

\newcommand{\eqn}[1]{(\ref{#1})}

 %%%%%%%%%%%%%%%%%%%%%%%%%%%%%%%%%%%%%%%%%%%%%%%%%%%%%%%%%%%%%%%%%%%%%%%%
 \def\beq{\be}
 \def\eeq{\ee}
 \def\bea{\ba}
 \def\eea{\ea}
 %%%%%
 \section{Introduction}
 T-duality, which in its simplest form states an equivalence between strings propagating on a circle of radius $R$ and those on a circle of inverse radius $\alpha'/R$,  is a
cornerstone  of the web of dualities that exist
within string theory and M-theory.
A natural question to ask is whether T-duality may be
generalised beyond the case of circular dimensions with $U(1)$ isometries
to strings whose target space contains non-Abelian
isometry groups.   In a pioneering work on the subject
\cite{delaossa:1992vc}  explains how to generalise
the procedure introduced by Buscher (for Abelian T-duality)
in \cite{Buscher:1987sk} .
Indeed,  the process of gauging isometries,
introducing Lagrange multipliers to enforce a flat
connection and integrating out the gauge fields
to produce a dual model,  was extended
to the case of non-Abelian isometries.
Other important foundational work on the subject 
includes \cite{Giveon:1993ai}-\cite{Alvarez:1994zr}.

\no
Beyond these initial breakthroughs two main difficulties emerged.
Firstly it seemed rather hard to obtain "interesting" dual backgrounds in
this manner and secondly the status of such non-abelian duality
transformations as full symmetries of string (genus) perturbation theory
is questionable \cite{Giveon:1993ai}, \cite{Alvarez:1993qi}.
Nonetheless, it is reasonable to consider the role of
non-abelian T-duality as a solution generating symmetry of the low energy
effective action of string theory, i.e. supergravity.   It is
of particular interest to address this question in the
context of the AdS/CFT correspondence  \cite{Maldacena:1997re}.

\no
A technical challenge that needed to be addressed, in light of
the AdS/CFT correspondence, was to understand non-abelian T-duality
in supergravity backgrounds with Ramond-Ramond fluxes.
This was first achieved in \cite{Sfetsos:2010uq} and has been
extended in a number of recent works in \cite{Lozano:2011kb} and \cite{Itsios:2012dc}. A brief review of
elementary aspects of non-Abelian T-duality can be found in \cite{Sfetsos:2011jw}.
Recently, a supersymmetric
solution of Type IIB containing an $AdS_6$ factor was  constructed using non-Abelian T-duality in \cite{Lozano:2012au}.

\no
In this letter, motivated by the AdS/CFT correspondence \cite{Maldacena:1997re}, we shall describe the
utility and application of non-Abelian T-duality to Type II supergravity
backgrounds with ${\cal N } = 1$ supersymmetry. We will find that
(up to subtleties to be discussed) backgrounds of the form presented in
\cite{Bah:2012dg} are found, starting from trademark solutions in Type IIB and non-abelian T-dualising them.
In particular, we will present two {\it new} solutions.
One  of the form $AdS_5\times S^2\times M_4$ in eleven-dimensional Supergravity and
another  in Massive IIA Supergravity that may be thought
as the 'cascading' version of the
first. Some subtle points will be discussed, but we leave a
detailed study of the properties of these geometries for \cite{Itsios:2013wd}.

\section{The Technique of Non-Abelian Duality}
In this letter we consider Type II backgrounds that have a freely acting $SU(2)$ symmetry such that the metric may be decomposed as
\ba
ds^2 &=& G_{\m\n}(x) dx^\m dx^\n + 2 G_{\m i}(x)  dx^\m  L^i + g_{ij}(x) L^i     L^j \ ,
\ea
where $\m=1,2,\dots , 7$ and $L^i$ are the left invariant Maurer--Cartan
forms $L^i = - i {\rm Tr}(g^{-1} d g)$. We also assume a similar ansatz for all the other fields.

\no
The  non-linear sigma model  corresponding to this background is
\be
S = \int d^2 \sigma Q_{\m\n} \del_+ X^\m \del_-X^\n
+  Q_{\m i} \del_+ X^\m L_-^i + Q_{i\m} L_+^i  \del_- X^\m + E_{ij} L_+^i L_-^j \ ,
\ee
where
\be
Q_{\m\n} = G_{\m\n} + B_{\m \n} \ ,\quad
Q_{\m i} = G_{\m i } + B_{\m i } \ , \quad  Q_{ i\m} = G_{ i\m } + B_{ i\m } \ ,
\quad E_{ij} = g_{ij} +  b_{ij}\ .
\ee
To obtain the dual sigma model one first gauges the isometry  by making the replacement
\be
\del_\pm  g \to D_\pm g = \del_\pm g - A_\pm g \ ,
\ee
 in the pulled-back Maurer--Cartan forms. In addition, a Lagrange multiplier term $-i {\rm Tr}(v F_{+-})$  is added to enforce a flat connection.

\no
After integrating this Lagrange multiplier term by parts, one can solve for the gauge fields to obtain the T-dual model. The final step of the process is to gauge fix the
redundancy, for instance, by setting $g= \mathbb{1}$.   In this way one obtains the Lagrangian,
\be
\tilde S  = \int d^2 \sigma \, Q_{\m\n}\del_+ X^\m \del_- X^\n + (\del_+ v_i + \del_+ X^\m Q_{\m i})( E_{ij} + f_{ij}{}^k v_k)^{-1} (\del_- v_j - Q_{j\m} \del_- X^\m )\ ,
\label{tdulal}
\ee
from which the T-dual metric and B-field can be ascertained.  As with Abelian T-duality the
dilaton receives a shift from performing the above manipulations in a path integral.

\no
In principle, one ought to repeat the above process in a formalism that caters for the full background including RR fluxes.
However the correct transformation rules for the RR fields may be obtained with the following recipe (which can be motivated
by for instance considering the pure spinor superstring as in \cite{Sfetsos:2010xa}   for the related case of fermionic T-duality).  One observes
that after T-duality, left and right movers naturally couple to two different sets of vielbeins for the dual geometry.
Since these two sets of frame fields describe the same metric they are related by a Lorentz transformation which we denote by  $\Lambda$.
This Lorentz transformation induces an action on spinors defined by the invariance property of gamma matrices:
\be
\Omega^{-1}  \Gamma^a  \Omega =  \Lambda^a{}_b \Gamma^b\ .
\label{spino1}
\ee
To find the dual RR fluxes one simply acts by multiplication  from the right with this $\Omega$ on the RR bispinor (or equivalently Clifford multiplication of the RR poly form).  More explicitly,  the T-dual fluxes $\hat{P}$ are given by
 \be
\hat{P} =  P \cdot \Omega^{-1} \ ,
\label{ppom}
\ee
where
\be
{\rm IIB}:\ P  = {e^{\Phi}\ov 2} \sum_{n=0}^4 \slashed{F}_{2n+1}\ ,%\ov (2n+1)!} \ ,
\qq
{\rm IIA}:\ \hat{P} ={e^{\hat \Phi}\ov 2} \sum_{n=0}^5 \slashed{\hat F}_{2n}\ . %\ov (2n)!} \ ,
\ee
The chirality of the theory is preserved/switched when the isometry group dualised has even/odd dimension respectively.
Full details and general expressions 
for the dual geometry, including alternate gauge fixing choices, 
will be reported in the forthcoming publication
\cite{Itsios:2013wd}.

\section{The conformal case: T-dual of the  Klebanov--Witten background}

In \cite{Klebanov:1998hh} the system of D3-branes at the tip of the conifold was studied.
The gauge theory on the branes is an $\cN=1$ superconformal
field theory with product gauge
group $SU(N)\times SU(N)$ and bifundamental  matter fields.
This  gauge theory is dual to the Type IIB  string theory
on $AdS^5 \times T^{(1,1)}$ with $N$ units of RR flux on the $T^{(1,1)}$.
The geometry and the $5$-form self-dual flux form, are given by
 \ba
 ds^2 &=&  \frac{r^2}{L^2} dx_{1,3}^2 + \frac{L^2}{r^2} dr^2 + L^2 ds^2_{T^{1,1}} \ ,
 \nonumber \\
 %g_s F_{(5)} &=& (1 + \ast) \frac{r^3}{L^4} dr \wedge dx_{1,3} \ .
 F_{(5)} &=& {4 \ov g_s L} \left({\rm Vol}(AdS_5) - L^5 {\rm Vol} (T_{1,1})\right)\ .
 \ea
Here $T^{(1,1)}$ is the homogenous space $(SU(2) \times SU(2))/U(1)$ with the diagonal embedding of the $U(1)$.
It has an Einstein metric with $R_{ij} = 4 g_{ij}$ given by
 \be
 ds^2_{T^{(1,1)}} =  \lambda_1^2 (d\th_1^2 + \sin^2 \th_1 d\phi_1^2) +   \lambda_2^2  ( d\th_2^2 + \sin^2 \th_2 d\phi_2^2) + \lambda^2 \left( d\psi + \cos \th_1 d \phi_1 + \cos \th_2 d\phi_2 \right)^2\ .
 \ee
 with $\lambda_1^2= \lambda_2^2 = \frac{1}{6} $ and $\lambda^2 = \frac{1}{9}$.   In these conventions $L^4 = \frac{27}{4} g_s N \pi$ ensures that the charge $\int_{T^{1,1}} F_5 = 16 \pi^4 N$ is correctly quantised for integer $N$.

\no
We perform a dualisation with respect to the $SU(2)$ isometry that acts on the  $\theta_2, \phi_2, \psi$ coordinates.  The result of the dualisation procedure\footnote{To obtain this we actually chose to fix the gauge symmetry by taking $\theta_2=\phi_2 = v_2 = 0$, rather than simply $g= \mathbb{1}$ since it makes manifest the residual isometries.  Additionally for aesthetic reasons we rename $v_1 = 2x_1$ and $v_3 = 2x_2$ and set $L=1$. } is a target space with NS fields given by
\ba
\widehat{ds}^2 &=& ds_{\rm AdS_5}^2 + \l_1^2 (d\theta_1^2  + \sin^2 \theta_1 d\phi_1^2 )+ {\l_2^2\l^2\ov \D} x_1^2 \s_{\hat 3}^2
   \nonumber \\
&&  + {1\ov \D} \left( (x_1^2 + \l^2 \l_2^2 )dx_1^2 + (x_2^2 + \l_2^4) dx_2^2 + 2 x_1 x_2 dx_1 dx_2   \right) \ ,
  \nonumber \\
\widehat{B} &=& - {\l^2\ov \D} \left[x_1 x_2 dx_1  + (x_2^2 + \l_2^4) dx_2 \right]\wedge \s_{\hat 3} \ ,
\label{T11dual}
\\
e^{-2 \widehat{\Phi}} &=& {8\ov g_s^2} \Delta \ ,
\nonumber
\ea
where $ \s_{\hat 3} = d\psi + \cos\th_1 d\phi_1$ and
\be
\Delta \equiv  \l_2^2 x_1^2 + \l^2 (x_2^2 + \l_2^4 ) \ .
\label{T11dualn}
\ee
This geometry is regular and the dilaton never blows up.
For a fixed value of $(x_1 , x_2)$ the remaining directions give a squashed three sphere.
The metric evidently has a $SU(2) \times U(1)_\psi$ isometry.

\no
Following the procedure outlined above---see \cite{Itsios:2013wd} for details---
one can determine the RR fluxes that support this geometry to be
\ba
\label{T11dualforms}
 \widehat{ F}_2 & = & \ {8 \sqrt{2}\ov g_s}\ \l_1^4\ \l\  \sin \theta_1 d\phi_1\wedge d\theta_1  \ ,
 \nonumber\\
 \widehat{F}_4 & = & - {8 \sqrt{2} \ov g_s} \l_1^2\ \l_2^2\ \l {x_1\ov \D}  \sin \theta_1 d\phi_1\wedge d\theta_1 \wedge d\psi
\wedge (\l_2^2 x_1\ dx_2- \l^2 x_2\ dx_1)\ .
\ea
This background enjoys ${\cal N}= 1$  supersymmetry and its explicit Killing spinors can be
determined by the expression
\be
\hat{\eta}  = \Omega \cdot \eta
\label{zzzxxx}\ee
where $\eta$ are the Killing spinors of the Klebanov--Witten background and the $\Omega$ matrix defined in (\ref{spino1}) has the form
\be
\Omega = {1\ov \sqrt{\D}}  \G_{11} \left(  -\l \l_2^2 \G_{123} +  \l_2 x_1 \G_1   + \l x_2  \G_3   \right) \  .
\ee
One could anticipate this result since the $U(1)_R$ symmetry
commutes with the $SU(2)$ used in the T-duality.
Hence one expects the corresponding isometry to be preserved after dualisation.
In the Appendix E of \cite{Itsios:2013wd} we have
explicitly checked the vanishing of the Killing spinor equation of type IIA for the T-dual background.
In this letter we follow a shortcut for preserving supersymmetry, first put forward in \cite{Sfetsos:2010uq}, by explicitly verifying
that the Killing spinors of the Klebanov--Witten
background have vanishing spinor-Lorentz-Lie derivative along the three Killing vectors that generate the $SU(2)$ isometry.
Indeed, let us give some details. For a Killing vector
$k$ this derivative is defined by
\be {\cal L}_k \eta = k^\mu D_\mu \eta +
\frac{1}{4}\nabla_\mu k_\nu \G^{\mu \nu} \eta \ .
\ee
Inserting the form of the Killing
vectors (for details see \cite{Itsios:2013wd}) we obtain that
\ba {\cal L}_{k^{(4)}}
\eta &=& 0 \ ,
\nonumber
\\ {\cal L}_{k^{(5)} } \eta &=& -\frac{1}{4} \sin(\phi_1) \csc(\th_1) \left(
\Gamma_{12} + \Gamma_{\hat{1}\hat{2}} \right) \eta \ ,
\label{killk2}
\\
{\cal L}_{k^{(6)} }\eta &=&
\frac{1}{4} \cos(\phi_1) \csc(\th_1) \left( \Gamma_{12} + \Gamma_{\hat{1}\hat{2}} \right)
\eta \ .
\nonumber
\ea
Thus, using that the Killing spinor $\eta$ in the background 
satisfies (see for instance \cite{Arean:2006nc}))
\begin{equation}
\Gamma_{12}\eta=- \Gamma_{\hat{1}\hat{2}}\eta= i\eta\ ,
\label{fgkj2}
\end{equation}
we find that the Killing spinor has vanishing Kosman derivative along the $SU(2)$.  This
corresponds to the statement that in the dual field theory the supersymmetry is not charged
under the $SU(2)$ flavour symmetries. Hence we anticipate that supersymmetry is preserved
after performing a T-duality along this $SU(2)$ with the Killing spinor
in the dual having the form (\ref{zzzxxx}).

\no
It is interesting to ask what are the charges of extended objects in this background.
Because of the non zero NS two-form, the Chern--Simons terms
play an important role and in general, the notion of charge that is quantised is the Page charge.
There is a natural two-cycle in the geometry, $\Sigma_2  = \{ \theta_1,\phi_1\} $,
over which the D6-brane charge can be measured by integrating $\widehat{F}_2$.
One finds that the D3-brane charge has been converted to D6-brane charge after dualisation.
One might anticipate that the presence of $\widehat{F}_4$ would indicate that there are also D4-brane charges.
We were not able to find a suitable four-cycle in the geometry over which to measure such a charge and believe that the activation of
$\widehat{F}_4$ is required only to solve the supergravity equations of motion.  However, the existence or not of a four-cycle, the determination of the possible periodicities of the $x_1,x_2$ coordinates and generally understanding better global issues, are two problems that require a dedicated future study.

\no
A natural question to ask is, what is the field theory dual to this geometry. As a first step one might wish to calculate the central charge,
which essentially is done by measuring the volume of the internal manifold.  A remarkable feature of non-Abelian T-duality is that this volume
is conserved in the following sense;
\be
e^{-2\Phi} \sqrt{\det g} \, \Delta_{F.P.} = e^{-2\hat{\Phi}}  \sqrt{ \det \hat {g}}\ ,
\ee
where $\Delta_{F.P.} $ is  the Fadeev--Popov determinant that arises from gauge fixing to obtain the dual sigma model.
That is to say all of the complexity of the metric cancels against that of the dilaton leaving a rather simple result.
As we will see in the non-conformal case this implies that the central charges match up to an RG scale independent multiplicative constant.
Such a relation was first shown for gauged WZW models in \cite{Bars:1991pt},
but it is valid in the context of non-Abelian duality as well.

\no
The lift to eleven dimensions (along the circle with coordinate $x_{\sharp} $) of the geometry we found in eq.(\ref{T11dual}), is given by
\ba
ds^2 &=& \D^{1/3} \left( ds^2_{AdS_5} +  \l_1^2 (\s_{\hat{1}}^2 + \s_{\hat{2}}^2) \right)  +  \D^{-2/3} \left[(x_1^2 + \l^2 \l_1^2 )dx_1^2 \right.
\nonumber \\
&& \left. \quad  + (x_2^2 + \l_1^4) dx_2^2 + 2x_1 x_2 dx_1 dx_2
+  \lambda^2 \lambda^2_1 x_1^2 \s_{\hat{3}}^2  +  \left( d x_{\sharp} + \frac{ \s_{\hat{3}} } {27} \right)^2 \right]\ ,
\label{ds11we}
\ea
where $\D$ is given in \eqn{T11dualn}.
The four-form flux field is given by
\be
F_4 = d( C_3 + B \wedge dx_{\sharp} ) = {1\ov 27} dx_2\wedge  \s_{\hat 1} \wedge \s_{\hat 2} \wedge \s_{\hat 3} + H \wedge dx_{\sharp}\ ,
\label{ds11wefl}
\ee
where $H=dB$ is computed using the expression for $B$ in \eqn{T11dual}.

\no
Recently, a class of ${\cal N}=1$ (generically non-Lagrangian) SCFT's found
as the IR  fixed point of the dynamics
of M5-branes wrapped on a  genus $g$ surface $\S_g$ was
engineered \cite{Bah:2012dg,Bah:2011vv}.
These field theories enjoy not only a $U(1)_R$ global symmetry but also an additional $U(1)$ global symmetry.
 Moreover in \cite{Bah:2012dg,Bah:2011vv} the geometrical dual to
these solutions was given.
 Rather remarkably our solution fits in this ansatz for the case of
genus zero (the sphere).  This is an intriguing connection
and certainly hints towards a field theoretic interpretation. However
two caveats must be made; firstly that the field theories of
\cite{Bah:2012dg,Bah:2011vv} are less well understood in general for
the case of genus zero and secondly that even within the
solutions presented in \cite{Bah:2012dg,Bah:2011vv},  ours is special.

\no
Our solution corresponds to a particular limiting value of the parameters that classify the eleven-dimensional solutions in  \cite{Bah:2012dg,Bah:2011vv}.
It is clear from the metric \eqn{ds11we} that our solution has a seven-dimensional submanifold given by the factor $AdS_5 \times S^2$.
One may dimensionally reduce such solutions along $x_1,x_2$, 
as well as $\psi$ and $x_\sharp$ and obtain a seven-dimensional effective theory
with $AdS_5 \times S^2$ as a vacuum. This might appear to be in contradiction with
the fact that $AdS_5 \times S^2$ is not a fixed point of the BPS equations of the corresponding
seven-dimensional gauged supergravity used in \cite{Bah:2012dg,Bah:2011vv, Maldacena:2000mw}.\footnote{In the notation of \cite{Bah:2012dg} our solution
corresponds to $\kappa =1$ and $|z|=1$ (see \cite{Itsios:2013wd} for details). For these values the BPS
system of eqs. (3.10) of \cite{Bah:2012dg} is not solved by real constant values for $g,\l_1,\l_2$.
This in turn implies that $AdS_5 \times S^2$ is not a solution.} The 
paradox is resolved by realizing that this effective theory
is not part of the seven-dimensional 
gauged supergravities used in \cite{Bah:2012dg,Bah:2011vv, Maldacena:2000mw}.
An analog
situation, with a reduction to seven dimensions leading to a  consistent action,
which however is not part of a previously known supergravity theory, 
was encountered in \cite{Itsios:2012dc}.

\no
Let us remark further on some similarity with the situation considered in \cite{Sfetsos:2010uq} where  the same
$SU(2)$ non-Abelian dualisation was performed on $AdS_5 \times S^5$.
In that case the resultant geometry corresponded to a
limit of the Gaiotto-Maldacena geometries \cite{Gaiotto:2009gz}, dual to ${\cal N}=2$ SCFTS presented in  \cite{Gaiotto:2009we}.  Although there supersymmetry was halved by the dualisation whereas here it is preserved, what we have here can be viewed as an ${\cal N}=1$ parallel to  \cite{Sfetsos:2010uq}.  Indeed, the theories considered in \cite{Bah:2012dg,Bah:2011vv} are really
${\cal N}=1$ cousins
of the Gaiotto  ${\cal N}=2$ theories and can be
obtained by integrating out some ${\cal N}=1$ scalars contained
in ${\cal N}=2$ vector multiplets.     An interesting question to ask is if one can use a similar procedure to dualise the entire flow between $AdS_5 \times S^5/\mathbb{Z}_2$ and $AdS_5 \times T^{1,1}$  geometries to provide a gravity description of the flow between the  ${\cal N}=2$ SCFTS in  \cite{Gaiotto:2009we} and the   ${\cal N}=1 $ in  \cite{Bah:2012dg,Bah:2011vv}.

\section{The non-conformal case: T-dual of the  Klebanov-Tseytlin solution}

Let us now turn our attention to non-conformal backgrounds obtained by placing $M$ fractional D3-branes i.e. D5-branes wrapping a contractible two cycle
of $T^{(1,1)}$ as in \cite{Klebanov:2000nc,Klebanov:2000hb}.
This modifies the field theory to be $SU(N) \times SU(N+M)$,
hence no longer conformal.  In fact this theory
has rich RG dynamics undergoing a sequence of Seiberg dualities
to lower rank gauge groups as one proceeds to the IR.
In the IR, strong coupling dynamics takes hold giving rise
to spontaneous $Z_{2M}$-symmetry breaking, confinement
and other non-perturbative effects.

\no
Let us here discuss the case of Klebanov-Tseytlin (KT) \cite{Klebanov:2000nc},
details of the full Klebanov Strassler
geometry \cite{Klebanov:2000hb}
and related ${\cal N}=1$ backgrounds \cite{varios}
appear in \cite{Itsios:2013wd}.

\no
The geometry is given  \cite{Klebanov:2000nc} by\footnote{The dilaton is constant
and we have set it equal to 1 so that there is no difference between string and Einstein frame.}
\be
  ds^2 = h^{-1/2}(r) dx_{1,3}^2 + h^{1/2}(r) \left(dr^2 + r^2 ds^2_{T^{1,1}} \right)
\label{ktxxx}\ee
where the warping function displays the characteristic 
logarithmic running
\be
 h = b_0 + \frac{P^2}{4 r^4} \ln (r/r_{\ast})\ .
\ee
This is supported by fluxes
\ba
 B = - T(r) \om_2 \ ,\qq F_3 = -P e^\psi \wedge \om_2\ ,  \qq F_5 =(1+ \ast ) K(r) \rm{vol}(T^{1,1})
\ea
where the forms $e^\psi$ and $ \om_2$ are the conventional ones defined on $T^{1,1}$ and may be found explicitly in  \cite{Klebanov:2000nc}.

\no
In fact, this is a particular solution of a class of
KT-geometries characterised by a set of functions
obeying some BPS equations.  Although in this letter we
only consider this special solution it can be shown that the whole ansatz can be non-Abelian
T-dualised and solves the supergravity equations of motion subject to the same BPS equations.

\no
Again we perform the non-Abelian duality with respect to an $SU(2)$ isometry and find a dual geometry given by
\ba
d\hat s ^2 &=&  h^{-1/2}(r) dx_{1,3}^2 + h^{1/2}(r)\left( dr^2 + \frac{r^2}{6}  (d\theta_1^2 + \sin^2 \theta_1 d\phi_1^2) \right)   +\widehat{ds}_3^2 \ ,
\\
d\hat s_3^2 &=& \frac{1}{2 r^2 \Delta   h^{1/2}(r)  } \left( 12 r^4 h(r) v_2^2 \sigma_{\hat{3}}^2  + 12 ( r^4 h(r) + 27 v_2^2    ) dv_2^2 +  9  (2 r^4 h(r) + {\cal V}^2  ) dv_3^2  + 108      {\cal V} v_2 dv_2 dv_3 \right)\ ,
\nonumber
\ea
with
\be
\Delta =  2 r^4 h(r) + {\cal V}^2  + 54 v_2^2   \ , \quad {\cal V}=   6 v_3 - T(r)  \ .
\ee
This geometry is supported in the NS sector by both a dilaton and a two-form,
\ba
\widehat{B} &=&  -\frac{T(r)}{ 6 \sqrt{2}}\sin \theta_1 d\theta_1 d\phi_1 + \frac{3 \sqrt{2}}{ \Delta} {\cal V} v_2 \sigma_{\hat{3}}\wedge dv_2 + \frac{1}{\sqrt{2} \Delta} (2 r^4 h(r) + {\cal V}^2  )             \sigma_{\hat{3}}\wedge dv_3 \ ,  \nonumber \\
e^{- 2\widehat{\Phi}} &=&  \frac{1}{81} r^2  h(r)^\ha  \Delta \ .
\ea
In the RR sector we find
\ba
   \widehat{F}_0  =  \frac{P}{9} \ , \qquad  \widehat{F}_2 =   \frac{2 K(r) - P{\cal V}  }{54\sqrt{2} } \sin \theta_1 d\theta_1 \wedge d\phi_1  + \frac{\sqrt{2} P {\cal V} v_2  }{ 3 \Delta} \sigma_{\hat{3}}\wedge dv_2 -  \frac{3 \sqrt{2} P v_2^2 }{ \Delta} \sigma_{\hat{3}}\wedge dv_3     \  \ , \nonumber \\
  \widehat{F}_4  =  \frac{v_2}{18 \Delta }  \sin \theta_1 d\theta_1 \wedge d\phi_1 \wedge d\psi \wedge \left(  - 9 ( 2 K(r) - P{\cal V}) v_2 dv_3   + 2  (P r^4 h(r) + {\cal V} K(r)  + 27 P v_2^2  ) dv_2  \right) \ . \nonumber \\
\ea

\no
The same argument for 
preserving supersymmetry for the case of the non-Abelian T-dual of the 
Klebanov--Witten background 
will also work in the case of the non-Abellian T-dual of the Klebanov--Tseytlin background.
The reason is that the Killing vectors of the original solution are still given by \eqn{killk2}
and also the projection by \eqn{fgkj2} (an additional projection 
along the radial direction just ensures the breaking of superconformal
invariance) \cite{Arean:2006nc}.

\no
The metric has some similarities with the case of the dualised KW theory which is to be expected.  However, the RR sector reveals a striking difference; this is  a solution of {\it{massive}} type IIA supergravity with the Roman's mass obeying a natural quantisation given by $P$ which measured the number of fractional branes prior to dualisation.   Indeed the Page charges of this solution,
\be
Q_{Page, D6}= \frac{1}{\sqrt{2}\pi^2}\int_{\theta\varphi} \widehat{F}_2  - \widehat{F}_0 \widehat{B}
=\frac{2Q}{27\pi}
\ , \qquad Q_{Page, D8}=\sqrt{2}\int \widehat{F}_0= \frac{\sqrt{2}P}{9}, \nonumber
\ee
show that what was D3 charge has become D6 charge and what was D5 charge has become D8 charge (a result which chimes well with the naive view of performing three T-dualities).  There is no obvious cycle for D4-brane charge to be measure over.
Before dualisation the duality cascade could be seen by studying  the charges.  Indeed, two
equivalent views \cite{Benini:2007gx}
of this are the changes seen in the D3 Maxwell charge as
the radial coordinate is varied or the jumps in the Page charge
induced by large gauge transformations such that $\frac{1}{4\pi^2} \int B_2$
changes by an integer. Indeed one finds an analogous behaviour in
the charges after dualisation again suggestive of some
field theory cascade interpretation.  One subtlety is that a change of $M$
units in the charges of the KT geometry becomes a change of $2M$
units in the transformed geometry. Giving a complete field theory
description of this set up remains an interesting problem,
but is beyond the scope of this letter;
see \cite{Itsios:2013wd}.

\no
As we indicated earlier the invariance of the stringy volume of the internal
manifold to be dualised,   has strong implications for the
central charge.  In particular if we calculate the central charge following
the procedure explained in \cite{Klebanov:2007ws} (modified slightly to accommodate
a dilaton that may depend on the internal dimensions) one finds in the original geometry of
eq.(\ref{ktxxx}),
\be
 c=\frac{2\pi^3}{27A'(r)^3} \
\ee
and after dualisation
\be
\widehat{c}  =  \frac{\sqrt{2} \pi^2 }{27A'(r)^3} \times  {\cal I}
\ee
where $A(r)$ is defined in  \cite{Klebanov:2007ws}  and given by
$e^{2A(r) } = h(r)^\frac{1}{3} r^\frac{10}{3}$ .
One sees that the two agree up to a single RG scale invariant constant
that is set by the periodicities of the dual coordinates.
More precisely this constant  ${\cal I}$  is determined entirely by the rather
subtle global properties of the T-dual coordinates, in this  case  we have
${\cal I} =  \int dv_3\int dv_2 \,  v_2$.
An important question for further study is to better understand such
global issues, either via the sigma model or via space time considerations.

 \section{Discussion}

Developing the field theory duals corresponding to these geometries
represents the most obvious open problem.  One approach is to consider
various D brane probes and 'define' the field theory
via its observables, calculated in a smooth background
(with all IR effects taken into account).
This analysis is
reported  in \cite{Itsios:2013wd}.
Nevertheless, a more canonical approach, based on
a careful field theory analysis following the lead of \cite{Gaiotto:2009we},
\cite{Benini:2009mz} may be in order.

\no
We believe that as well as developing the particular
cases studied above this work opens up many possible new lines of research.  Firstly a more general classification of
massive type IIA backgrounds that display similar signatures of cascade would be highly desirable.
Equally one could hope to use the techniques outlined above to find new and interesting classes of backgrounds.
Indeed in this work and other recent studies, it seems that we have only just started seeing the utility of these duality transformations.
In principle whenever a space time admits a non-abelian isometry these techniques might be applicable. There are,
of course, many such examples and we hope that further study will prove fruitful.

\section{Acknowledgements}
We would like to thank Francesco Benini, Eoin Colgain, 
Tim Hollowood, Prem Kumar,
Yolanda Lozano, Alberto Mariotti, Diego Rodriguez-Gomez, 
Brian Wecht  and especially to Kostas
Siampos for interesting discussion and correspondence.
The research of G. Itsios has been co-financed by the ESF and Greek
national funds through the Operational Program "Education and Lifelong Learning" of
the NSRF - Research Funding Program: ``Heracleitus II. Investing in knowledge
in society through the European Social Fund''.
This research is implemented (K.S.) under the "ARISTEIA" action
of the "operational programme education and lifelong learning" and is co-funded by
the European Social Fund (ESF) and National Resources. Daniel Thompson is supported in part by the Belgian Federal Science Policy
Office through the Interuniversity Attraction Pole P7/37, and in part by the
"FWO-Vlaanderen" through the project G.0114.10N and through an "FWO-Vlaanderen"
postdoctoral fellowship project number 1.2.D12.12N.

 %
 %
 %%%%%%%%%%%%%%%%%%%%%%%%%%%
 %
 %

\providecommand{\href}[2]{#2}\begingroup\raggedright\endgroup


\begin{thebibliography}{10}


\bibitem{delaossa:1992vc}
X.C. de~la Ossa and F.~Quevedo, {\it Duality symmetries from non-Abelian
  isometries in string theory},  Nucl. Phys. {\bf B403} (1993) 377, \href{http://arxiv.org/abs/hep-th/9210021}{{\tt hep-th/9210021}}.

\bibitem{Buscher:1987sk}
T.H. Buscher, {\it {A Symmetry of the String Background Field Equations}},
   Phys. Lett. {\bf B194} (1987) 59 and {\it {Path Integral Derivation of Quantum Duality in Nonlinear
  Sigma Models}},  Phys. Lett. {\bf B201} (1988) 466.
  %CITATION = PHLTA,B194,59;%%

\bibitem{Giveon:1993ai}
A.~Giveon and M.~Rocek, {\it {On nonAbelian duality}},  Nucl. Phys. {\bf
  B421} (1994) 173, \href{http://xxx.lanl.gov/abs/hep-th/9308154}{{\tt
  hep-th/9308154}}.
 %%CITATION = NUPHA,B421,173;%%


\bibitem{Curtright:1994be}
T.~Curtright and C.K. Zachos, {\it {Currents, charges, and canonical structure
  of pseudodual chiral models}},  Phys. Rev. {\bf D49} (1994) 5408,
  \href{http://xxx.lanl.gov/abs/hep-th/9401006}{{\tt hep-th/9401006}}.


% \cite{Sfetsos:1994vz}
\bibitem{Sfetsos:1994vz}
  K.~Sfetsos,
 {\it Gauged WZW models and non-Abelian duality},
  Phys. Rev. {\bf D50} (1994) 2784,
  \href{http://arxiv.org/abs/hep-th/9402031}{{\tt hep-th/9402031}}.
  %%CITATION = PHRVA,D50,2784;%%


\bibitem{Alvarez:1994zr}
E.~Alvarez, L.~Alvarez-Gaume, and Y.~Lozano, {\it {On non-Abelian duality}},
  Nucl. Phys. {\bf B424} (1994) 155,
  \href{http://xxx.lanl.gov/abs/hep-th/9403155}{{\tt hep-th/9403155}}.


%\cite{Alvarez:1993qi}
\bibitem{Alvarez:1993qi}
 E.~Alvarez, L.~Alvarez-Gaume, J.~L.~F.~Barbon and Y.~Lozano,
  {\it {``Some global aspects of duality in string theory}},
Nucl.\ Phys.\ B {\bf 415} (1994) 71
\href{http://xxx.lanl.gov/abs/hep-th/9309039}{{\tt hep-th/9309039}}.


%\cite{Maldacena:1997re}
\bibitem{Maldacena:1997re}
  J.~M.~Maldacena,
 {\it  The Large N limit of superconformal field theories and supergravity},
  Adv.\ Theor.\ Math.\ Phys.\  {\bf 2} (1998) 231,
 \href{http://arxiv.org/abs/hep-th/9711200}{{\tt hep-th/9711200}}.
  %%CITATION = HEP-TH/9711200;%%


%\cite{Sfetsos:2010uq}
\bibitem{Sfetsos:2010uq}
  K.~Sfetsos and D.~C.~Thompson,
  {\it On non-abelian T-dual geometries with Ramond fluxes},
  Nucl.\ Phys. {\bf B846} (2011) 21,
  \href{http://arxiv.org/abs/1012.1320}{{\tt arXiv:1012.1320}}.
  %%CITATION = ARXIV:1012.1320;%%



%\cite{Lozano:2011kb}
\bibitem{Lozano:2011kb}
  Y.~Lozano, E.~O.~Colgain, K.~Sfetsos and D.~C.~Thompson,
{\it Non-abelian T-duality, Ramond Fields and Coset Geometries},
  JHEP {\bf 1106} (2011) 106,
  \href{http://arxiv.org/abs/1104.5196}{{\tt arXiv:1104.5196}}.
  %%CITATION = ARXIV:1104.5196;%%




  %\cite{Itsios:2012dc}
\bibitem{Itsios:2012dc}
  G.~Itsios, Y.~Lozano, E.~O.~Colgain and K.~Sfetsos,
  {\it Non-Abelian T-duality and consistent truncations in type-II supergravity},
  JHEP {\bf 1208} (2012) 132,
  \href{http://arxiv.org/abs/1205.2274}{{\tt arXiv:1205.2274}}.
  %%CITATION = ARXIV:1205.2274;%%


 %\cite{Sfetsos:2011jw}
\bibitem{Sfetsos:2011jw}
  K.~Sfetsos,
  {\it Recent developments in non-Abelian T-duality in string theory},
  Fortsch.\ Phys.\  {\bf 59} (2011) 1149,
  \href{http://arxiv.org/abs/1105.0537}{{\tt arXiv:1105.0537}}.
  %%CITATION = ARXIV:1105.0537;%%

%\cite{Lozano:2012au}
\bibitem{Lozano:2012au}
  Y.~Lozano, E.~O.~Colgain, D.~Rodriguez-Gomez and K.~Sfetsos,
  {\it New Supersymmetric $AdS_6$ via T-duality},
    \href{http://arxiv.org/abs/1212.1043}{{\tt arXiv:1212.1043}}.


\bibitem{Bah:2012dg}
  I.~Bah, C.~Beem, N.~Bobev and B.~Wecht,
  {\it Four-Dimensional SCFTs from M5-Branes},
  JHEP {\bf 1206} (2012) 005,
  \href{http://arxiv.org/abs/1203.0303}{{\tt arXiv:1203.0303}}.
  %%CITATION = ARXIV:1203.0303;%%




%\cite{Itsios:2013wd}
\bibitem{Itsios:2013wd}
  G.~Itsios, C.~Nunez, K.~Sfetsos and D.~C.~Thompson,
  {\it Non-Abelian T-duality and the AdS/CFT correspondence:new $\cN=1$ backgrounds},
  \href{http://arxiv.org/abs/1301.6755}{{\tt arXiv:1301.6755}}.
  %%CITATION = ARXIV:1301.6755;%%


  %\cite{Sfetsos:2010xa}
\bibitem{Sfetsos:2010xa}
  K.~Sfetsos, K.~Siampos and D.~C.~Thompson,
  {\it Canonical pure spinor (Fermionic) T-duality},
  Class.\ Quant.\ Grav.\  {\bf 28} (2011) 055010,
  \href{http://arxiv.org/abs/1007.5142}{{\tt arXiv:1007.5142}}.
  %%CITATION = ARXIV:1007.5142;%%

%\cite{Klebanov:1998hh}
\bibitem{Klebanov:1998hh}
  I.~R.~Klebanov and E.~Witten,
  {\it Superconformal field theory on three-branes at a Calabi-Yau singularity},
  Nucl. Phys. {\bf B536} (1998) 199,
  \href{http://arxiv.org/abs/hep-th/9807080}{{\tt hep-th/9807080}}.
  %%CITATION = HEP-TH/9807080;%%


%\cite{Arean:2006nc}
\bibitem{Arean:2006nc}
  D.~Arean,
  {\it Killing spinors of some supergravity solutions},
   \href{http://arxiv.org/abs/hep-th/0605286}{{\tt hep-th/0605286}}.
  %%CITATION = HEP-TH/0605286;%%


%\cite{Bars:1991pt}
\bibitem{Bars:1991pt}
 I.~Bars and K.~Sfetsos,
 {\it Generalized duality and singular strings in higher dimensions},
 Mod. Phys. Lett. {\bf A7} (1992) 1091,
 \href{http://xxx.lanl.gov/abs/hep-th/9110054}{{\tt hep-th/9110054}}.
 %%CITATION = HEP-TH/9110054;%%

\bibitem{Bah:2011vv}
  I.~Bah, C.~Beem, N.~Bobev and B.~Wecht,
  {\it AdS/CFT Dual Pairs from M5-Branes on Riemann Surfaces},
  Phys. Rev. {\bf D85} (2012) 121901,
  \href{http://arxiv.org/abs/1112.5487}{{\tt arXiv:1112.5487}}.
  %%CITATION = ARXIV:1112.5487;%%


%\cite{Maldacena:2000mw}
\bibitem{Maldacena:2000mw}
  J.~M.~Maldacena and C.~Nunez,
  {\it Supergravity description of field theories on curved manifolds and a no go theorem},
  Int. J. Mod. Phys. {\bf A16} (2001) 822,
  \href{http://arxiv.org/abs/hep-th/0007018}{{\tt hep-th/0007018}}.
  %%CITATION = HEP-TH/0007018;%%


\bibitem{Gaiotto:2009gz}
  D.~Gaiotto and J.~Maldacena,
  {\it The gravity duals of ${\cal N}=2$ superconformal field theories},
 \href{http://arxiv.org/abs/0904.4466}{{\tt arXiv:0904.4466}}.
  %%CITATION = ARXIV:0904.4466;%%


\bibitem{Gaiotto:2009we}
  D.~Gaiotto,
  {\it $\cN=2$ dualities},  JHEP {\bf 1208} (2012) 034,
   \href{http://arxiv.org/abs/0904.2715}{{\tt arXiv:0904.2715}}.
%  %%CITATION = ARXIV:0904.2715;%%






%\cite{Klebanov:2000nc}
\bibitem{Klebanov:2000nc}
  I.~R.~Klebanov and A.~A.~Tseytlin,
  {\it Gravity duals of supersymmetric SU(N) x SU(N+M) gauge theories},
  Nucl. Phys. {\bf B578} (2000) 123,
  \href{http://arxiv.org/abs/hep-th/0002159}{{\tt hep-th/0002159}}.
  %%CITATION = HEP-TH/0002159;%%


 %\cite{Klebanov:2000hb}
\bibitem{Klebanov:2000hb}
  I.~R.~Klebanov and M.~J.~Strassler,
  {\it Supergravity and a confining gauge theory: Duality cascades and chi SB resolution of naked singularities},
  JHEP {\bf 0008} (2000) 052,
  \href{http://arxiv.org/abs/hep-th/0007191}{{\tt hep-th/0007191}}.
  %%CITATION = HEP-TH/0007191;%%

\bibitem{varios}
%\cite{Butti:2004pk}
%\bibitem{Butti:2004pk}
  A.~Butti, M.~Grana, R.~Minasian, M.~Petrini and A.~Zaffaroni,
  {\it The Baryonic branch of Klebanov--Strassler solution: A supersymmetric family of $SU(3)$ structure backgrounds},\hfill\break
  JHEP {\bf 0503} (2005) 069,
  \href{http://arxiv.org/abs/hep-th/0412187}{{\tt hep-th/0412187}}.\\
  %%CITATION = HEP-TH/0412187;%%
%\cite{Conde:2011aa}
%\bibitem{Conde:2011aa}
  E.~Conde, J.~Gaillard, C.~Nunez, M.~Piai and A.~V.~Ramallo,
  {\it A Tale of Two Cascades: Higgsing and Seiberg-Duality Cascades from type IIB String Theory},\hfill\break
  JHEP {\bf 1202} (2012) 145,
  \href{http://arxiv.org/abs/1112.3350}{{\tt arXiv:1112.3350}}.\\
  %%CITATION = ARXIV:1112.3350;%%
%\cite{Gaillard:2010qg}
%\bibitem{Gaillard:2010qg}
  J.~Gaillard, D.~Martelli, C.~Nunez and I.~Papadimitriou,
  {\it The warped, resolved, deformed conifold gets flavoured},
  Nucl.\ Phys. {\bf B843} (2011) 1,
  \href{http://arxiv.org/abs/1004.4638}{{\tt arXiv:1004.4638}}.
  %%CITATION = ARXIV:1004.4638;%%

%\cite{Benini:2007gx}
\bibitem{Benini:2007gx}
  F.~Benini, F.~Canoura, S.~Cremonesi, C.~Nunez and A.~V.~Ramallo,
  {\it Backreacting flavors in the Klebanov-Strassler background},
  JHEP {\bf 0709} (2007) 109,
  \href{http://arxiv.org/abs/0706.1238}{{\tt arXiv:0706.1238}}.
  %%CITATION = ARXIV:0706.1238;%%

%\cite{Klebanov:2007ws}
\bibitem{Klebanov:2007ws}
  I.~R.~Klebanov, D.~Kutasov and A.~Murugan,
  {\it Entanglement as a probe of confinement},
  Nucl. Phys. {\bf B796} (2008) 274,
  \href{http://arxiv.org/abs/0709.2140}{{\tt arXiv:0709.2140}}.
  %%CITATION = ARXIV:0709.2140;%%

%\cite{Benini:2009mz}
\bibitem{Benini:2009mz}
  F.~Benini, Y.~Tachikawa and B.~Wecht,
  {\it Sicilian gauge theories and N=1 dualities},
  JHEP {\bf 1001} (2010) 088,
  \href{http://arxiv.org/abs/0909.1327}{{\tt arXiv:0909.1327}}.
  %%CITATION = ARXIV:0909.1327;%%


\end{thebibliography}
\end{document}